\documentstyle[preprint,aps,prd]{revtex}
\def \MSbar {\vbox{\hrule\kern 1pt\hbox{\rm MS}}}
\def \GeV { {\ \rm GeV} }

\begin{document}
%
\rightline{PSU/TH/164}
\bigskip
\centerline{
The Limiting Curve of Leading Particles from Hadron-Nucleus Collisions
at Infinite A}
\bigskip
\centerline{ A. Berera$^{\dagger}$ \footnote{\it Current address:
Department of Physics and Astronomy, Vanderbilt University,
Nashville, Tennessee 37235}, M. Strikman$^{\dagger}$, W. S. Toothacker$^{\ddagger}$, W. D. Walker$^{**}$, and J. J. Whitmore$^{\dagger}$ }
\smallskip
\centerline{\it
$^{\dagger}$Department of Physics,
Pennsylvania State University,
University Park, PA 16802}
\vskip 0.1cm
\centerline{\it
$^{\ddagger}$Department of Physics,
Pennsylvania State University,
Mont Alto, PA 17237}
\vskip 0.1cm
\centerline{\it
$^{**}$Department of Physics,
Duke University,
Durham, NC 27706
}

\begin{abstract}
We argue that
as the atomic number of the target nucleus $A \rightarrow \infty$,
the multiplicity
of leading particles in hadron-nucleus collisions tends to a 
finite
limit.  The limiting multiplicities for
various particle production are computed for both proton
and pion projectiles.  Signatures at finite A
are discussed. Data from 100$\GeV/ c$ central hadron-nucleus 
collisions are analyzed and found to be in
qualitative agreement with this picture.
\end{abstract}

\medskip


\bigskip

In Press Physics Letters B 1997

\bigskip
\bigskip
\bigskip
\eject

%

The production of leading  hadrons in high-energy hadron-nucleus scattering
has been studied for many years
(for a recent review, see \cite{busza}). The inclusive data
indicate a strong suppression of leading particles   in central
hadron-nucleus (h-A) collisions.  Even for heavy nuclei, the
main contribution to the inclusive spectra comes
from the interactions near the nuclear edge and not
the center. More recent attempts to enhance the 
contribution of collisions at small impact parameters by selecting events with
multiple nucleon emission \cite{walker} confirm that the  leading 
hadron multiplicity rapidly decreases with increasing A.

There are few theoretical results for the leading particle multiplicity
from central h-A collisions.  The earliest
 phenomenological results were based on the applications 
of various models of soft strong interactions, notably 
 Gribov reggeon calculus.
  There are some clear problems with applying this approach to
the projectile fragmentation region (which we will discuss below)
and in fact very few explicit treatments of the leading particle
multiplicity problem have been made.  An alternative approach is to apply
perturbative QCD (pQCD), in which parton correlators are the elementary
building blocks, see e.g. \cite{geiger}. 
 There have been theoretical arguments \cite{mueller},
and more recently experimental evidence from nuclear size dependence
tests \cite{ryan}, which show that  hadronization 
 of high-energy partons  occurs outside the
nucleus.
 
 The new observation of this paper is that for large
nuclei, and a sufficiently high energy projectile, the origin of
leading particles may have a very simple explanation based on
the underlying parton dynamics and may be calculable through the pQCD
ingredients of the strong interaction dynamics.

Before addressing our proposed pQCD mechanism, let us first review
the expectations of the Gribov-Glauber theory of 
hadron-nucleus interactions \cite{Gribov69}.
This theory is known to describe  quantitatively the total and elastic
cross sections of high-energy hadron-nucleus scattering, 
see \cite{Jenmil} and references therein. Corresponding
reggeon calculus diagrams include multipomeron exchanges with a
hadron, which are expressed through the vertices where the  hadron 
couples to $n$ pomerons. These vertices involve complicated hadron intermediate
states and hence are not known independently.  The use of
 the Abramovskii, Gribov, Kancheli (AGK) cutting rules \cite{AGK}
allows one  to describe also the production of 
  hadrons in the central rapidity
range, for a review see e.g. \cite{BS}.
However, the  production of particles in the projectile  fragmentation
region requires additional information because the AGK cutting rules
are not applicable for this kinematic region \cite{AGK}. 
A generic Gribov-Glauber approach does not have predictive power 
here.
One has to know the structure of
the vertex (so called Mandelstam cross) for coupling of the hadron with
 $n$ pomerons,
including information on how the energy of the projectile is split between
several pomerons. One example of such a model is an  eikonal type model
\cite{shab} where it is assumed that the energy is split equally between 
$n$
interactions. Another is 
the additive quark model
where the eikonal is applied to the interactions of constituent 
quarks 
\cite{ASSh}.
One common feature of these models  is that in the
 limit of $A\rightarrow \infty$ the leading particle multiplicity tends
to zero. 

At the same time according to QCD-parton model concepts, leading
particles originate from the emerging fast partons
of the collision debris.  Here a leading 
or fast parton is defined as one carrying
a large fraction of the projectile's longitudinal momentum.
In particular there is a wide
rapidity separation between a fast
parton and the sea partons.
A reasonable criterion is that
$ x > 0.2$, where $x$ is its momentum fraction. 
In addition,  the absolute momentum of the fast parton should be sufficiently
large so that it 
cannot easily 
interact softly with
the medium. 
According to the space-time picture of strong interactions
\cite{gribov}
a parton fluctuates into other hadronic
states in a time governed by the uncertainty principle.
Thus a parton of three-momentum $p$ fluctuates into a state of mass $m$
 in a time,
\begin{equation}
t = \frac{2p}{m^2}.
\label{tcond}
\end{equation}
The state of lowest mass gives the characteristic time required
for the point-like parton to become spatially extent.
Conservatively taking $m \sim m_{\rho}$ (versus $m_{\pi}$)
one finds that the condition,
\begin{equation}
t =\frac{2p}{m^2}> 2R_{A}
\label{long}
\end{equation}
is satisfied at A=200 for $p \geq 15$ GeV/c, where $R_A$ is the radius of a
nucleus of atomic number $A$.
 This requires projectiles with energy $E > 75$ GeV.

In the ideal parton model limit,
such fast partons would interact rarely with the surrounding nuclear medium.
In this ideal description,
the interaction of the hadron projectile with the nuclear target would be
primarily through the wee partons in the former.  As such, 
a fast parton or a coherent configuration of fast partons
would filter through essentially unaltered.  This would
imply that the leading particle spectrum for a given hadronic projectile
on any nuclear target would be the same \cite{kan}.  Such universality
in the spectra is qualitatively inconsistent with experiment \cite{busza}.

QCD also predicts that fast partons will
lose a negligible fraction
of
their longitudinal momentum 
if $p_{inc} \rightarrow \infty$ and $A=const$.
This expectation for fast partons to retain their
longitudinal momentum can be tested from the 
nuclear size dependence of the Drell-Yan cross section.  Recall
that Drell-Yan production depends kinematically only  on the 
longitudinal momentum fractions of the two impinging partons.
Thus if there is no substantial attenuation of longitudinal momentum
for a fast parton, the Drell-Yan cross section will scale linearly with
A.
This is what experiment finds 
\cite{drellyan}.

In addition, QCD
predicts \cite{bod} that fast partons undergo 
quasi-elastic rescatterings
which give them a transverse momentum
$<p_T^2>$ that increases linearly with the
distance traveled in the nuclear medium.  Up to
geometrical corrections the dependence goes approximately
as $A^{1/3}$.  
This QCD prediction has been confirmed by the Drell-Yan experiment 
\cite{drellyan}.
The $p_T$ - broadening in the dimuon spectrum for
hadron projectile $h_p$ and target nucleus $A$,
$ \Delta \langle p_T^2\rangle_{h_pA}$,
is given by
the transverse momentum of
the initiating partons in the projectile.
The effect of coherence between collisions of the fast
parton with the nuclear media
does not change the linear dependence on distance for the average 
$\langle p^2_T \rangle_{h_pA}$,
as discussed recently in \cite{Baier}.

In pQCD transverse broadening and energy loss are related as
\cite{Baier}\footnote{This relation depends weakly  on the approximations
used in \cite{Baier}, see discussion in \cite{Dok}.}
\begin{equation}
\Delta E \approx {\alpha_s N_c \over 8} \Delta \langle p_T^2 \rangle L/2,
\label{relation}
\end{equation}
where $L$ is the path in the nuclear matter.
Transverse momentum smearing implies that
a type of limiting curve
is expected.  This limiting behavior arises because
an increase in the relative transverse momentum between two
leading partons
decreases the probability for two such partons to coalesce.
In particular one expects a high coalescence probability
for relative transverse momenta
$p_{T_R} \equiv \sqrt{2 \langle p_T^2 \rangle}$
that are of typical
hadronic scale, e.g.
$p_{T_R} < p_{T_R}^{typical} \sim \sqrt{2(0.3)^2}  ~\GeV/c \approx 0.420 ~\GeV/c$.
Increasing the nuclear volume, thus increasing $p_{T_R}$,
decreases the likelihood of coalescence at least 
as $\propto {1 \over p_T^4}$ for large $p_T$, see equation (8) below. So 
 in the limit
of infinite nuclear volume, $A \rightarrow \infty$, the coalescence
probability goes to zero, and the leading partons should
fragment independently and so produce independent jets.
At the same time the fractional energy loss tends to zero for a fixed large
value of $A$.
Thus
the $z$-distribution of these jets will not depend on A, although
the transverse momentum will increase with A.
Obviously one can chose the double limit of $E \propto A^k$,
$ A \rightarrow \infty$
so that  $p_T^2 \ge cA^{n}$ while 
both condition (\ref{long}) and  the inequality 
${\Delta E \over E_h} \rightarrow 0$ are satisfied.
This corresponds to 
\begin{equation}
k> n+1/3.
\end{equation}

In this limit, when the leading partons
(mostly quarks in a nucleon projectile or $q$ and ${\bar q}$ in
a pion) fragment independently,
it is possible to calculate the leading parton
production cross
section integrated over the transverse momentum $p_T$.
Let $zd\sigma_A^{h/h_p}(z)/dz$ denote the one-particle inclusive 
differential cross
section, from collisions of a hadronic projectile
$h_p$ on a nucleus A, for the production of a hadron $h$ that
carries a longitudinal momentum
fraction $z$, integrated over its transverse momentum.
The differential leading particle multiplicity is then defined as
\begin{equation}
z\frac{dN^{h/h_p}_A(z)}{dz} \equiv z\frac{1}{\sigma^{inel}_{h_p}(A)}
\frac{d\sigma_A^{h/h_p}(z)}{dz},
\end{equation}
where $\sigma^{inel}_{h_p}(A)$ is the inelastic cross section
for $h_p - A$ scattering.
In the limit of interest, $zdN^{h/h_p}_A(z)/ dz$ takes
on the asymptotic form:
\begin{equation}
z\frac{dN_A^{h/h_p}(z)}{dz}=\sum_{a=q, {\bar q}, g} \int_z^1 dx 
\frac{z}{x}D^{h/a}(\frac{z}{x},Q^2)
f_{a/h_p}(x,Q^2),
\label{leadmult}
\end{equation}
involving the convolution of
$f_{a/h_p}(x,Q^2)$, the distribution of parton $a$ in the hadron 
projectile $h_p$, and
$D^{h/a}(z,Q^2)$, the fragmentation function for parton $a$ into
hadron $h$.  Eq. (\ref{leadmult})
leads to a steep decrease of the hadron spectrum
at large $z$ because at $Q^2 \sim 1 ~\GeV^2$ the structure functions
drop as
$xf_{q/h_p}(x, Q^2) \sim (1-x)^n$ with $n \sim 2-3$ and
the dominant fragmentation functions $zD^{\pi/q}(z, Q^2) \sim (1-z)$ based on
quark counting rules or at most are constant in low virtuality
approximations \cite{kaidalov,berger}.
The hard interactions that transfer transverse momenta to
the fast partons are at low virtuality.
We set the virtuality at $Q^2=1 ~\GeV^2$.

For all our calculations,
we used the parton distributions
of GRV \cite{gluck1} since
they are determined down to the low-virtuality range
$Q^2 \sim 1 \GeV^2$.  Two types of fragmentation functions
were used.  One was determined by the EMC collaboration \cite{emc}
at $<Q^2> = 20 \GeV^2$ and the other was calculated
\cite{kaidalov} for low virtuality from the quark-gluon
string model (QGSM) \cite{kaidalov}.
The EMC fragmentation functions set a lower bound on
our multiplicity estimates, since scaling violations
will enhance fragmentation at low virtuality and large $z$.
As a cross check, we fit the EMC data for $D^{{\pi^+}/u}(z)$
to the higher twist form calculated in  \cite{berger}.
We then evolved this down to $Q^2=1 ~\GeV^2$.  In this region
in particular the Berger form also becomes
nonzero at $z=1$ which is consistent with the QGSM form.
Numerically we found that the fitted form
of \cite{berger} was a factor of one
to two times bigger than the QGSM form.
We consider this discrepancy acceptable.

All three fragmentation
functions suppress light quark fragmentation into
protons.  The EMC data indicate a factor of 4 to 8 depletion
of protons versus $\pi^+$ in the largest multiplicity region
of leading particles $z \sim 0.2 - 0.3$.  The higher
twist mechanism in \cite{berger} is inapplicable for proton fragmentation.
The QGSM results \cite{kaidalov} give zero fragmentation
into protons. 

The limiting curve for proton (solid) and pion (dashed) projectiles
are shown in Fig. 1 using the QGSM fragmentation functions.
One typically
finds the leading behavior as $z \rightarrow 1$ to
be $zdN_{\infty}^{h/h_p} (z)/ dz \sim (1-z)^{\alpha_{h/h_p}}$.
Here $\alpha_{h/h_p}$ will be referred to as the leading
exponent for produced species $h$ from projectile $h_p$
with $\alpha_{T/h_p}$ denoting the exponent for the total leading
particle multiplicity.  For the proton projectile with QGSM fragmentation
functions, we find
$\alpha_{{\pi^+}/p} \sim 4.4$, $\alpha_{\pi^-/p} \sim 5.3$ and
$\alpha_{T/p} \sim 4.5$.  
For the EMC fragmentation functions $\alpha_{X/p} \sim 6$ for all species
$X$ except for the proton, where $\alpha_{p/p} \sim 7$,
and $\alpha_{T/p} \sim 6$.
For the pion projectile with QGSM fragmentation functions, we find
$\alpha_{{\pi^+}/{\pi^+}} \sim 1.6$, $\alpha_{\pi^-/\pi^+} \sim 2.8$ and
$\alpha_{T/\pi^+} \sim 1.8$;
for the EMC fragmentation functions
$\alpha_{T/\pi^+} \sim 3.0$.
In all cases, multiplicity distributions
for ${\bar p}$ and $\pi^-$ projectiles are obtained by applying
charge conjugation to those for the $p$ and $\pi^+$ projectiles, respectively.

Let us now turn to what we might expect for the leading
particle multiplicities.  In a large nucleus, 
say $A=200$,
the relative transverse momentum that two leading partons acquire
can be estimated following \cite{gavin} through 
the $p_T$ - broadening of the Drell-Yan 
spectrum. Using data \cite{drellyan} for 
$ \Delta \langle p_T^2\rangle_{p A  \sim 200} \approx 0.114 ~\GeV^2$
 and  the average number of struck nucleons in the Drell-Yan events 
${\bar n}_{(A = 200)} = 5.3$ \cite{fs1}:
\begin{equation}
p_{T_R~A} \approx 2 \sqrt{ \Delta \langle p_T^2\rangle_{p \ A}{\bar n}_{A}/
({\bar n}_{A}-1)}|_{A \sim 200} \approx 
 0.75 ~\GeV/c.
\label{p75}
\end{equation}
This number should be considered as a lower limit since in
QCD $p^2_T$ - broadening is larger for the interaction of partons with smaller
virtualities \cite{LQS} and in our case the 
virtuality is much smaller than in the Drell-Yan
processes studied in \cite{drellyan} 
for $\langle M^2_{\mu^+\mu^-} \rangle=30 ~\GeV^2$.
Within the $z>0.2$ region, we estimate the coalescence
probability into a pion, $P^{\pi}_c(p_T^2)$, for a
$q {\bar q}$ pair with relative
momentum $p_{T_R}$.  It is easy to show in the constituent quark model that
\begin{equation}
P_c^{\pi}(p^2_{T_R}) = |F_{\pi}(p^2_{T_R})|^2
\label{Ak}
\end{equation}
where $F_{\pi}(p^2)$ is the pion form factor.
Using the Vector-Dominance-Model fit for
$F_{\pi}(p^2)$,
one finds for Eq. (\ref{p75}) $P_c^{\pi}(0.5) \approx 0.25$.
Further suppression comes from the presence of gluons in the pion wave function which on average carry half of the pion momentum.

For large A, the above mechanism for
leading particle
production goes as $A^{-2/3}$. One finds a similar A-dependence  
for the fragmentation of a diquark to a nucleon.
Another mechanism for the
production of leading particles which is similar to \cite{BH} is  
the filtering of color singlet small transverse size clusters of 
the projectile. This is analogous to
the propagation of ultrarelativistic positronium through
say a piece of lead.
Using equations derived in \cite{FS91}, one finds
for large A that this mechanism leads 
to a multiplicity of leading pions (protons)
$\propto A^{-2/3}(A^{-4/3})$.

Another ``background" to our mechanism is the coherent 
diffractive production of leading particles. Its contribution to $zdN^{h/h_p}_A/dz \propto A^{-1/3}$
for $A \rightarrow \infty$ and somewhat slower for $A=12 - 200$ \cite{fms}.

Although we see no expectation for the asymptotic form in Eq. (\ref{leadmult})
even for $A=200$, we feel
large nuclei may still inhibit coalescence. To test this claim,
since the leading correction to the asymptotic behavior 
is due to diffractive events,
it is preferable to perform data analyses which would
explicitly exclude diffractive events by selecting the events 
where the nucleus breaks up.
For this
these asymptotic expectations can be compared with 
data from central hadron-nucleus collisions
from Fermilab experiment E597
\cite{tooth,walker,whit}.

The data were obtained using the Fermilab 30-inch hybrid bubble chamber 
spectrometer with associated downstream particle identifiers (DPI). The bubble 
chamber, in a 2T magnetic field, provided a visual target and vertex detector 
as well as a spectrometer for the slower produced particles. The faster 
particles were momentum analyzed using the fringe field of the bubble chamber 
magnet and seven planes of proportional wire chambers
and three drift chambers. In addition to 
being filled with liquid hydrogen, the bubble chamber also contained thin 
nuclear foils of Mg, Ag and Au. The beams consisted of 100 
GeV/c ${\bar p}, p, \pi^+$ and $\pi^-$. 
Mass identification for these particles was 
provided by {\u C}erenkov counters in the beam-line. Additional experimental details 
are given elsewhere \cite{whit}.

To make the relevant comparisons, leading particle 
spectra have been obtained by
combining $p/{\bar p}$ and combining 
$\pi^{\pm}$ projectile data. To study 
the nuclear thickness dependence of the data, use is made of the 
known correlation between the number of nuclear collisions and the number of 
observed slow protons \cite{ander}. The number 
($n_p$) of slow protons (with momentum less than 1.3 GeV/c) is determined by 
performing a visual ionization scan of each interaction occurring in
the nuclear foil targets.
The selection of $n_p \ge 1$ automatically removes all diffractive events.

In Fig. 2a 
the differential multiplicity
$zdN_A^{\pi/\pi}(z)/dz$ is  shown
for the data from 
$\pi^- A \rightarrow h^+$ (combined with $\pi^+A \rightarrow h^-$)
with $n_p$ =  1 or 2.  In these figures, $h^+(h^-)$ indicates
a positively (negatively) charged hadron, mostly pions.
The dashed curve is a fit to the form $(1-z)^\alpha$
and demonstrates that this form is an acceptable representation
of the data with $\alpha= 3.27 \pm 0.48$.
Figs. 2b and 2c show the resulting values 
of $\alpha$ as a function of $n_p$ for the $\pi$ projectile to
$\pi^+/\pi^-$ and $p/{\bar p}$, respectively, with 
$z > 0.2$ and for produced hadrons
having the same or opposite charge 
to that of the beam projectile. 
The solid (dashed) horizontal lines at the right side 
of the figure are 
the theoretical
predictions using the QGSM fragmentation functions in Fig. 2b and 
the EMC fragmentation functions in Fig. 2c
for the opposite (like) charge leading particle exponent.

We next compare various integrated multiplicities from
the  experiment
to the predictions.
Let $I_{h/h_p}(z_m)$ denote
the integrated multiplicity of produced hadrons
of type
$h$ with $z>z_m$ from a projectile $h_p$ with $h$ replaced
by $T$ when all the produced hadrons are summed.  For the
predictions, these are
understood to be for $A \rightarrow \infty$.
In table 1 the limiting predictions for the integrated
multiplicities are compared with the experimental
results for $n_p \geq 15$.
Fig. 2d 
shows the E597 data for the
integrated multiplicity with $z_m = 0.2$ as a function of $n_p$.
The solid (dashed) horizontal lines at the right are the
predictions from the QGSM fragmentation functions
for $p$ ($\pi$) projectiles. \footnote{ Using QGSM versus GRV
proton distribution functions \cite{QGSMDF}, the predicted multiplicity is
about the same.  All the leading exponents are about two smaller, which
is consistent with the slower decrease of the former
as $x \rightarrow 1$.} 
The data seem to be consistent with the idea of limiting multiplicity
with the magnitude as determined by the model.

It is also interesting to examine multiplicity
ratios.  Denote the ratio as,
\begin{equation}
R^{{h_1}/{h_2}}_{h_p}(z)
 \equiv \frac{I_{h_1/h_p}(z)}{I_{h_2/h_p}(z)}.
\end{equation}
For the ratio between protons and pions, only the EMC fragmentation
functions are available.  The calculated values are
$R^{{p+{\bar p}}/{\pi^+ +\pi^-}}_p(0.2)=0.18$ and
$R^{{p+{\bar p}}/{\pi^+ +\pi^-}}_p(0.3)=0.17$.
To examine the depletion of leading protons from a proton
projectile, the lower bound estimates from the EMC fragmentation
functions are 
$I_{p/p}(0.2)=0.009$ and $I_{p/p}(0.3)=0.002$.
The statistics on experimental data for $p \rightarrow p$
are insufficient to quote.
It is worth noting that the forward multiplicity
for large $n_p$ is substantially larger than
in eikonal models where usually it is assumed that the energy is
equally  divided between 
all exchanges, see e.g. \cite{shab}. In these models one should
expect that almost no particles are produced for $z \ge 0.2$ for 
the case in which  more than 5 nucleons have been struck.

In our estimate of $p_T$ - broadening we have used experimental data on 
Drell -Yan production at 400 GeV/c. No estimates of the energy dependence of 
the $p_T$ - broadening are available at the moment. 
The model used in the   analysis 
 \cite{Baier} leads to energy independent
energy losses. However this model
assumes that the soft cut-off does not depend on energy.
It seems natural to expect a certain increase of 
hardness of the soft interaction with energy leading to an 
increase of 
$p_T$ - broadening  with an increase of energy. 
So fewer leading particles will
be produced in the central collisions as the energy of $hA (AA)$ collisions 
increases from
the SPS energy range to the RHIC or LHC  energy range
both because of further suppression of the coalescence effect and the scaling
violations for the quark fragmentation functions.

An implication of the current analysis is that in high energy
central $AA$ collisions very few baryons should be left in the fragmentation
region.
In fact our analysis indicates that the probability for a baryon to carry
more than
$z = 0.2 $ of the projectile's momentum fraction is less than $1 \%$.
To conserve baryon number,
a guess is that nucleons should move on average at least 4 units of
rapidity to the central rapidity region for RHIC and beyond.

M.S. thanks A.Mueller for discussions on the subject over the last 6  years.
We would also like to thank Yu.Dokshitzer for discussion of the
energy losses in QCD.
Financial support was provided in part by
the U. S. Department of Energy
and by the U. S. National Science Foundation.

\eject
FIGURE CAPTIONS

Figure 1: Limiting curves for leading $\pi^{\pm}$ production
from $p$ - $A$ (solid) and $\pi^+$ -$A$ (dashed) collisions.

Figure 2: Experimental results for $h-A$ collisions 
at 100$GeV/c$: (a) Differential multiplicity
for $\pi^- A \rightarrow \ \pi^+$ combined with 
$\pi^+A \rightarrow \ \pi^-$ for events with $n_p=1,2$ ($n_p$ is the
 number of slow protons). The dashed curve is a fit to the form
$(1-z)^{\alpha}$; 
(b) Leading exponent $\alpha$  for
$\pi A \rightarrow \pi$, solid circles (open boxes) when the
 produced $\pi$ has the same (opposite) charge to that of the beam projectile;
 the horizontal lines
on the right are the theoretical limits  from QGSM
fragmentation functions to the same  (dashed) and opposite (solid)
charge leading particle; (c) Leading exponent $\alpha$ 
for $\pi A \rightarrow \ p ({\bar p})$,
the horizontal lines
on the right are the theoretical results from EMC
fragmentation functions to like (dashed) and opposite (solid)
charge leading particle; 
(d) The integrated multiplicity of hadrons with 
$z_m > 0.2$ for $\pi^{\pm}$ (solid circles)
and proton (open boxes) beam projectiles as a function of
$n_p$. The horizontal lines on the right of the figure are the
 theoretical asymptotic limits for $\pi$ (dashed) and proton (solid)
 projectiles.
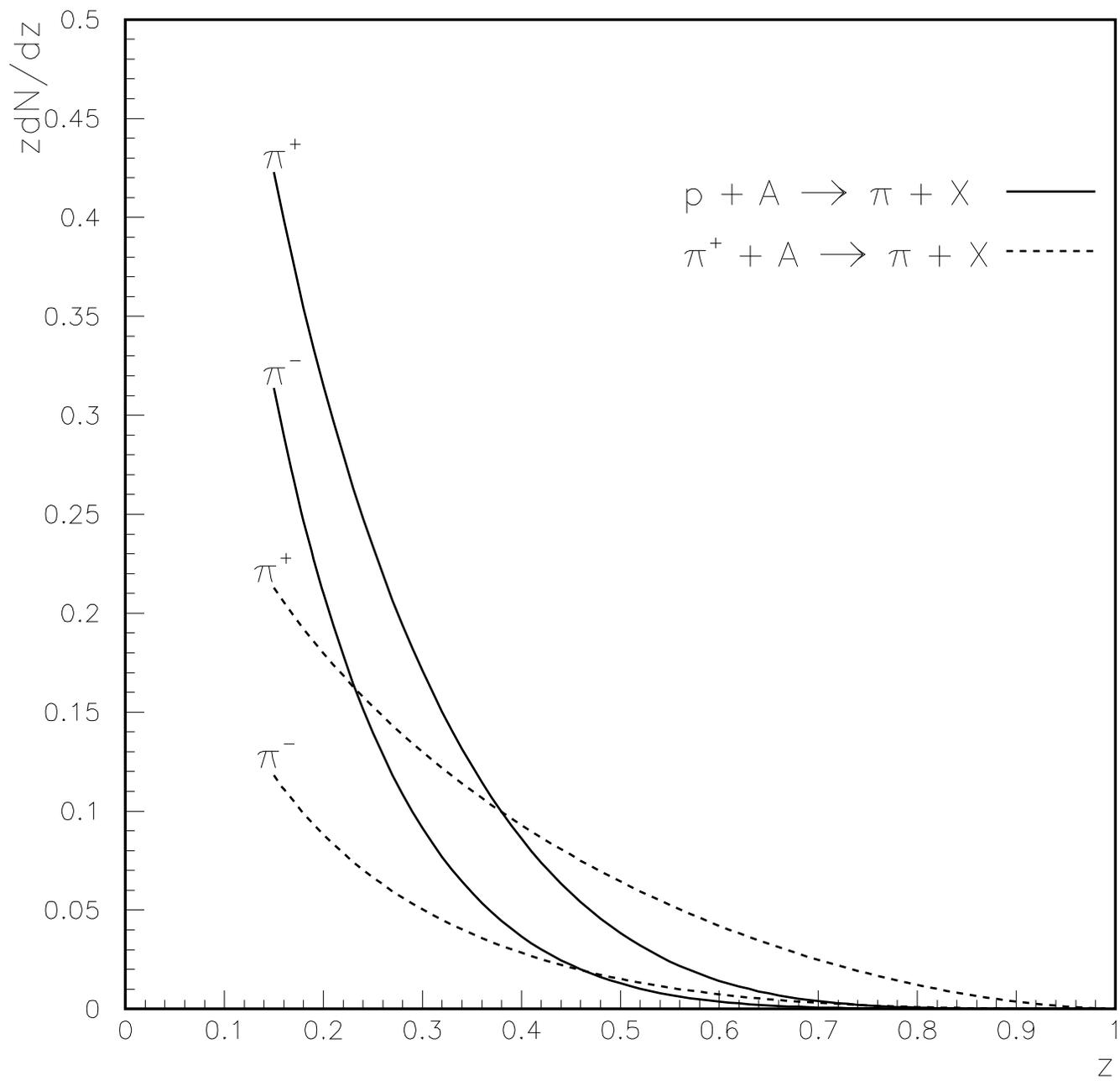
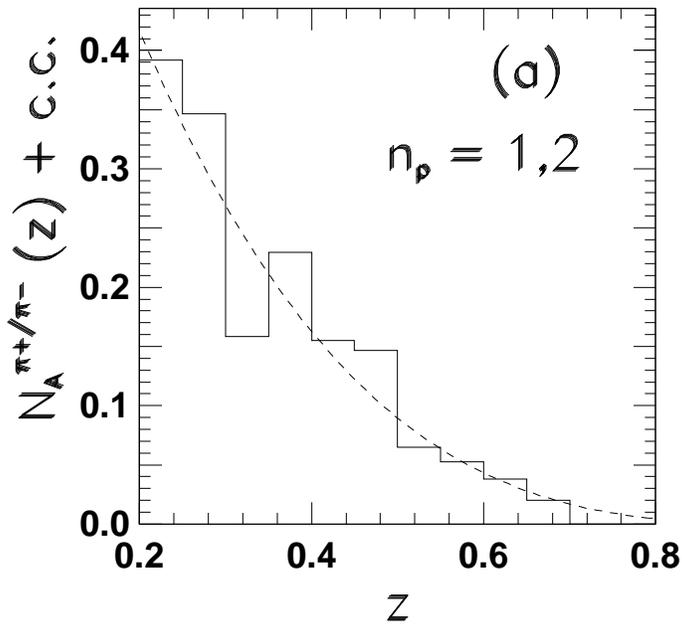
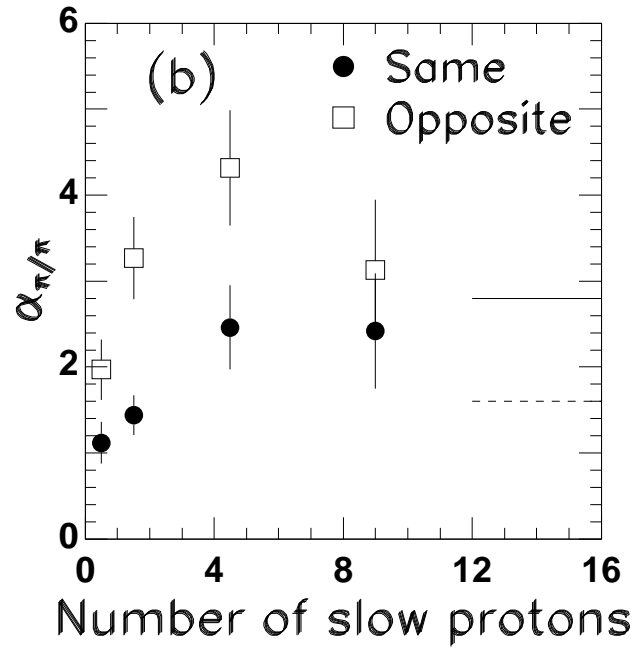
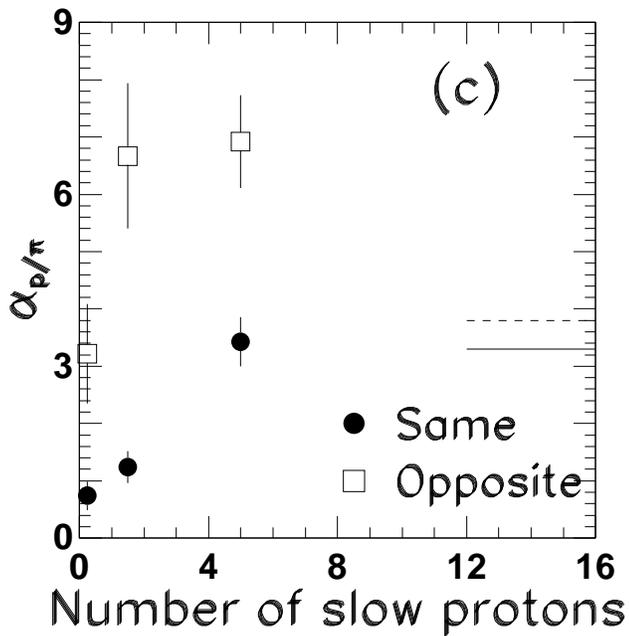
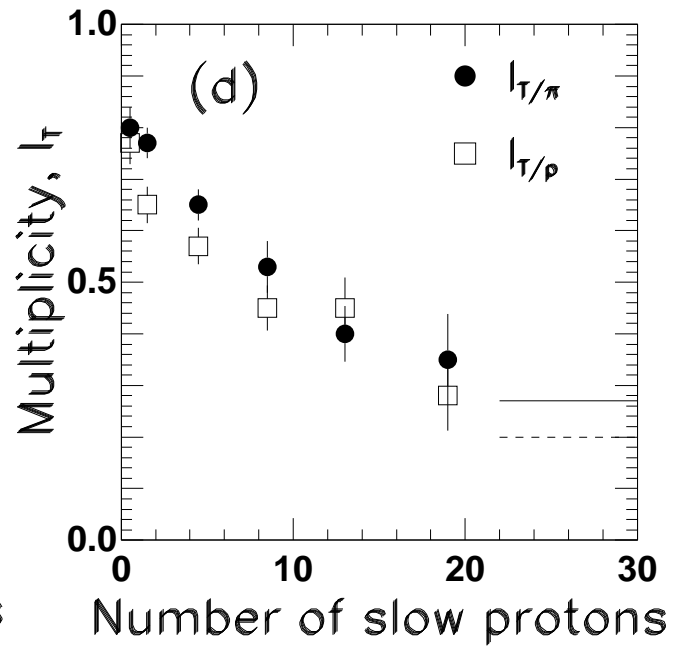
\end{document}